\documentclass[fleqn,10pt]{wlscirep}
\usepackage[utf8]{inputenc}
\usepackage[T1]{fontenc}
\usepackage{url,graphicx,epsfig,graphics,subfigure,psfrag,amsmath,amssymb,epstopdf,enumerate,lineno,hyperref}
\usepackage{diagbox}
\usepackage{xcolor}
\usepackage{pythonhighlight}
\usepackage{amssymb}
\usepackage{enum erate}
\usepackage{bm}
\usepackage{array}
\usepackage{multirow}
\usepackage{wrapfig}
\usepackage{fullpage}
\usepackage{overpic}
\usepackage{verbatim}
\usepackage{soul}
\usepackage[normalem]{ulem}

\title{Motion-driven quantum dissipation in an open electronic system with nonlocal interaction}

\author[1,*]{Feiyi Liu}

\author[1]{Min Guo}

\author[1,**]{Mingyang Liu}

\author[2]{Ruanjing Zhang}

\author[3]{Yang Wang}

\affil[1]{School of Physics, Electrical and Energy Engineering, Chuxiong Normal University, Chuxiong, 675000, China}

\affil[2]{Institute of Theoretical Physics, School of Science, Henan University of Technology, Zhengzhou, 450001, China}

\affil[3]{School of Big Data and Basic Science, Shandong Institute of Petroleum and Chemical Technology, Dongying, 257061, China}

\affil[*]{fyliu@cxtc.edu.cn}
\affil[**]{liumingyang@cxtc.edu.cn}

\begin{abstract}
In this paper, we study excitations and dissipation in two infinite parallel metallic plates undergoing relative motion. The degrees of freedom of the electrons in both plates are modeled using the 1+2 dimensional Dirac field, and a nonlocal potential is selected to describe the interaction between the two plates. The internal relative motion is introduced via a Galilean boost, with one plate assumed to slide relative to the other. We then calculate the effective action of the system and derive the vacuum occupation number in momentum space using a perturbative method. Numerical plots reveal that the vacuum occupation number, as a function of momentum, is isotropic for a motion speed $v = 0$ and anisotropic for nonzero $v$.
The relative motion induces energy transfer between the plates, leading to on-shell excitations in a manner analogous to the dissipative process of the Schwinger effect. Consequently, we study the motion-induced dissipation effects and the dissipative forces through the quantum action. Numerical results demonstrate that both the imaginary part of the quantum action due to the motion boost and the dissipative force exhibit a threshold as functions of $v$, and both are positively correlated with $v$.
\end{abstract}

\begin{document}




\flushbottom
\maketitle

\thispagestyle{empty}

\section{Introduction}
The study of open quantum systems (OQS) has gained significant attention over the decades in the field of condensed matter physics~\cite{Xu1474579}.
An important feature of OQSs is dissipation and its related dynamics that arise when the systems are coupled with environments.
For such OQSs, one can partially trace over the degrees of freedom (DOFs) of the environments to obtain the effective dynamics of the system \cite{Joichi101143,FEYNMAN2000547,doiS0129055X23500150,PhysRevE.75.031107,Jin101063}. This is similar to the process occurring in models of effective field theory, where one can obtain the effective action by integrating over some irrelevant DOFs~\cite{DEGRANDE201321,Wen101093,vasiliev1998functional,PhysRevD.58.123508,Peskin101063}.  However, there exists special cases of OQSs that do not obviously couple to environments. For instance, consider two parallel plates placed close together, where the internal DOFs of the plates couple with each other via some effective potentials. An external force acts on one of the plates to keep them moving uniformly, and this relative motion induces dissipation and a frictional force within the plates~\cite{jentschura2016friction}.

Dissipation refers to the irreversible transfer of energy. In other words, energy is carried by particles or quasi-particles~\cite{Wang2022MPLB}. For instance, in the dynamical Casimir effect involving two metal plates separated by a vacuum gap, on-shell photons are excited in the vacuum gap between the plates~\cite{PhysRevA.100.032510}. If the relative speed between the two plates is small, on-shell photons cannot be excited; however, the internal DOFs within the plates are still excited~\cite{PhysRevD.84.025011}. Another example in condensed matter physics is that two insulated plates stick together and slide relative to each other, causing phonons at the interfaces to be excited due to the sliding contact~\cite{Duan101115}. The quantum spins in the two plates interact with each other through magnetic exchange or dipole coupling, and this interaction causes the relative motion to excite magnons in the plates, also representing this type of energy transfer~\cite{5257437}. Electronic friction is also a prime example, in this case two metal plates stick together and move relative to each other, with the electrons in each plate interacting through direct exchange coupling~\cite{PERSSON2000145,Dou101063}. The models discussed above are all semiclassical and do not strictly account for quantum effects.

In a detailed discussion of OQS, the influence of quantum effects on microscopic systems cannot be ignored, especially the dissipation described by the quantum action. 
A functional approach has been tried to study the quantum effective dynamics of moving, planar, dispersive mirrors and etc, in different numbers of dimensions~\cite{PhysRevD.76.085007}.
Viotti \textit{et al} studied the dissipative effects and decoherence induced on a particle moving at constant speed in front of a dielectric plate in quantum vacuum, using a Closed-Time-Path (CTP) integral formulation to account for finite-temperature corrections~\cite{PhysRevD.99.105005}.
Farias \textit{et al} discussed the Casimir friction phenomenon in a system consisting of two flat, infinite, and parallel graphene sheets coupled to the vacuum electromagnetic (EM) field~\cite{PhysRevD.95.065012}.
Farías \textit{et al}  considered a particle moving in front of a dielectric plate, and studied the dissipation and the decoherence of the particle’s internal DOFs~\cite{PhysRevD.93.065035}. 
In previous work, we focused on the quantum motion-driven dissipation effect in a plasmon-magnon coupling system, and calculated the plasmon production probability and the sliding frictional force in the system~\cite{wang2024magnon}.
These studies attempt to explain quantum effects, especially dissipation, in OQSs from various perspectives, which has inspired our work.

In this paper, we study the quantum excitations and dissipation of two infinite parallel metallic plates moving relative to each other along their interface. This system is similar to the 3D topological insulators, such as $Bi_2Se_3$, $Sb_2Te_3$ and $Bi_2Te_3$~\cite{AndoJPSJ2013,PhysRevB.84.165120}. Massive Dirac electrons are considered in this model and a nonlocal Anderson mixed coupling is introduced to transmit friction, instead of relying on vacuum fluctuations of the electromagnetic field.
We assume a relative sliding is driven by an external field for excitations and dissipation, and analyze the interaction of electrons in the two plates via a nonlocal potential.
To study the excitations, we start with the DOFs associated with each plate using massive Dirac fermions and write the effective action of the electrons by considering one plate sliding relative to the other. By focusing on the time evolution of the system, we calculate the probability of particle production and derive the vacuum occupation number of the fixed plate perturbatively. Since the process of relative motion inducing on-shell excitations is very similar to the Schwinger effect, we can study the quantum effects of this process through the method of path integrals, and then obtain the imaginary part of the quantum action and the dissipative force. 

The main structure of this paper is as follows. In Section \ref{model}, the microscopic model will be given. Section \ref{particle_production} discuss the time evolution of the system and calculatethe probability of particle production. In Section \ref{excitations_L}, we focus on the excitations of the system. Section \ref{dissipation} is about quantum effects and dissipation. Section \ref{conclusion} is the conclusion of
this work.

\section{The microscopic model}\label{model}
\begin{figure}
  \centering
  \includegraphics[width=7cm]{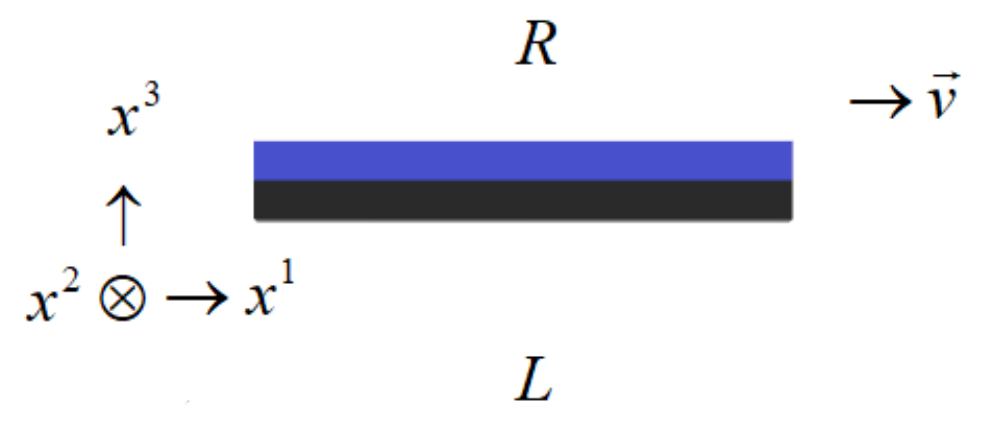}
  \caption{Schematic picture of the system.}
  \label{fig_schematic}
\end{figure}

\indent The model of system we consider in this study is the interface between two infinite metallic plates, which is shown in Fig.~\ref{fig_schematic}. For easy identification, one plate is marked as L-plate and the other as R-plate. 
Since our purpose is to study motion driven excitations and dissipation, relative sliding is required in the system. Here we assume that the L-plate is at rest in the laboratory frame, and the R-plate is moving along the interface. 
The total area of each plate is $V$, and the total evolution time for fiction of the system is $T$. To analyze the model, a 1+3 dimensional space-time coordinate system is constructed as $X=\begin{pmatrix}x^0, x^1, x^2, x^3\end{pmatrix}^\mathsf{T}$, where $x^0=ct$ is the time coordinate with the speed of light $c$ and $\{x^1,x^2,x^3\}$ are the Cartesian coordinates. The axis of $x^3$ is set to be perpendicular to the L-plate. The space-time coordinates for the 1+2 dimenstional space-time domain of the L-plate read $x=\begin{pmatrix}x^0, x^1, x^2\end{pmatrix}^\mathsf{T}$ and the spacial components are written as $\emph{\textbf{x}}=\begin{pmatrix}x^1, x^2\end{pmatrix}^\mathsf{T}$. We follow the convention of natural unit such that $\hbar=c=1$. The corresponding covariant components of a space time vector is $k_\mu=\eta_{\mu\nu}k^\nu$, i.e. $k_0=k^0$, $k_1=-k^1$ and $k_2=-k^2$, where $\eta_{\mu\nu}=\mathrm{diag}(1,-1,-1,-1)$ is the Minkovski metric. So the inner product of two space time vectors is $k\cdot x=k^0x^0-\emph{\textbf{k}} \cdot \emph{\textbf{x}}$.

To study the excitations of large numbers of materials, we start from the DOFs living on each plates via the 1+3 dimensional massive Dirac fermions, which can be defined as:
\begin{align}
&\psi_{\mathrm{R/L}}(x)=\begin{pmatrix}c_{\mathrm{R/L}\uparrow}(x), c_{\mathrm{R/L}\downarrow}(x)\end{pmatrix}^\mathsf{T}, \\
&\psi^{\dagger}_{\mathrm{R/L}}(x)=\begin{pmatrix}c^{\dagger}_{\mathrm{R/L}\uparrow}(x), c^{\dagger}_{\mathrm{R/L}\downarrow}(x)\end{pmatrix}.
\end{align}
Here $c^{\dagger}_{\mathrm{R/L}\uparrow}(x)$ and $c^{\dagger}_{\mathrm{R/L}\downarrow}(x)$ represent the components of the Dirac spinor, since surface states of both topological insulators are fermions. 
The free parts of the action corresponding to the free Dirac fermions on the two plates of our system can be expressed as~\cite{Golkar2014}: 
\begin{align}\nonumber
S^0=&\int \mathrm{d}^3x\psi_\mathrm{L}^{\dagger} [\mathrm{i}\partial_0 - v_\mathrm{F} \sigma^1 (-\mathrm{i}\partial_1)-v_\mathrm{F} \sigma^2 (-\mathrm{i}\partial_2)-\sigma^3v_\mathrm{F}m]\psi_\mathrm{L}\nonumber\\&
+\int \mathrm{d}^3x\psi_\mathrm{R}^{\dagger} [\mathrm{i}\partial_0 - v_\mathrm{F} \sigma^1 (-\mathrm{i}\partial_1)-v_\mathrm{F} \sigma^2 (-\mathrm{i}\partial_2)-\sigma^3v_\mathrm{F}m]\psi_\mathrm{R}.
\label{action}
\end{align}
The upper term describes the free action of the electrons living on the surface of L-plate, and the second line is for the R-plate. 
The integral measure refers to $\int \mathrm{d}^3x'=\int_{-T/2}^{T/2} \mathrm{d}x'^0 \int_{-\infty}^{\infty} \mathrm{d}x'^1 \int_{-\infty}^{\infty} \mathrm{d}x'^2 $. $v_\mathrm{F}$ is the Fermi velocity of the fermions, and
$\sigma$ is the Pauli matrix as:
\begin{align}
\sigma^1=\begin{pmatrix}0&1\\1&0\end{pmatrix},\sigma^2=\begin{pmatrix}0&-\mathrm{i}\\ \mathrm{i}&0\end{pmatrix},\sigma^3=\begin{pmatrix}1&0\\0&-1\end{pmatrix}.
\end{align}
Usually, real experiments take into account the influence of an external magnetic field, which causes the system to exhibit spin-orbit coupling, resulting in a mass term in the action. For example, an uniform applied magnetic field $\emph{\textbf{B}}=\begin{pmatrix}0, 0, v_\mathrm{F}m\end{pmatrix}^\mathrm{T}$ is directed along the $x^3$ axis.

In this study, electrons on the two plates are considered to couple via a direct-exchange interaction, similar to the Anderson mixed model~\cite{PhysRevLett.119.046001}. It involves local interactions and the electron-electron interaction typically involves direct exchange, ignoring nonlocal electron exchange coupling. However, the interaction occurs between two many-electron systems with nonlocal coupling, it should include not only a direct exchange term but also a nonlocal term. By taking a real-valued nonlocal potential $U(x,y)=U(x^0,\emph{\textbf{x}};y^0,\emph{\textbf{y}})$, this nonlocal Anderson mixed model where the fermions belonging to the two plates interact with each other can be written as:
\begin{align}
S^{\mathrm{int}} = g \int \mathrm{d}^3x \int \mathrm{d}^3y [\psi^{\dagger}_\mathrm{L}(x) U(x,y) \psi_\mathrm{R}(y) +h.c.].
\end{align}
Here $g$ is constant of small coupling, and $h.c.$ means the hermite conjugate of the term $\psi^{\dagger}_\mathrm{L}(x) U(x,y) \psi_\mathrm{R}(y)$. 
Since the nonlocal potential $U$ indicates the interaction propagates at a finite speed and the Anderson mixed coupling is used to describe quantum electrodynamics (QED), the most natural expression for $U$ is the Feynman propagator of a massless particle, i.e., a Green's function  with poles $k^0=\pm |\bm{k}|$, as:
\begin{align}
U(x^0,\emph{\textbf{x}};y^0,\emph{\textbf{y}})=\int \frac{\mathrm{d}^3k}{(2\pi)^3} \frac{\exp[\mathrm{i}k^0(x^0-y^0)-\mathrm{i}\emph{\textbf{k}} \cdot (\emph{\textbf{x}}-\emph{\textbf{y}})]}{\sqrt{(k^0)^2-\emph{\textbf{k}}^2+\mathrm{i}\epsilon}}.
\label{potential}
\end{align}
 The integral measure in the momentum space refers to $\int \frac{\mathrm{d}^3k}{(2\pi)^3}=\int_{-\infty}^{\infty} \frac{\mathrm{d}k^0}{2\pi} \int_{-\infty}^{\infty} \frac{\mathrm{d}k^1}{2\pi}\int_{-\infty}^{\infty} \frac{\mathrm{d}k^2}{2\pi}$. By introducing the (1+2) dimensional Fourier momentum $k=\begin{pmatrix}k^0, k^1, k^2\end{pmatrix}^\mathrm{T}$, the spacial components can be written as $\emph{\textbf{k}}=\begin{pmatrix}k^1, k^2\end{pmatrix}^\mathrm{T}$.
In Eq. \eqref{potential}, it obviously has two singular point on the $k^0$ complex plane, which means $U$ can be regarded as a propagator of massless particles. 

Now, we can obtain the total action of the system as:
\begin{align}\nonumber
S=&S^0+S^{\mathrm{int}}\nonumber\\
=&\int \mathrm{d}^3x\psi_\mathrm{L}^{\dagger} [\mathrm{i}\partial_0 - v_\mathrm{F} \sigma^1 (-\mathrm{i}\partial_1)-v_\mathrm{F} \sigma^2 (-\mathrm{i}\partial_2)-\sigma^3v_\mathrm{F}m]\psi_\mathrm{L}\nonumber\\&
+\int \mathrm{d}^3x\psi_\mathrm{R}^{\dagger} [\mathrm{i}\partial_0 - v_\mathrm{F} \sigma^1 (-\mathrm{i}\partial_1)-v_\mathrm{F} \sigma^2 (-\mathrm{i}\partial_2)-\sigma^3v_\mathrm{F}m]\psi_\mathrm{R}\nonumber\\&
+g \int \mathrm{d}^3x \int d^3y [\psi^{\dagger}_\mathrm{L}(x) U(x,y) \psi_\mathrm{R}(y) +h.c.].
\end{align}
Since our goal is to study motion-dependent dissipation, for simplicity, we assume that the L-plate is at rest in the laboratory frame, while the R-plate moves at a constant velocity, $\emph{\textbf{v}}=(0,v)^\mathsf{T}$, along the $x^1$-axis due to an external force.
This movement results in a Galilean boost acting on the R-plate along the $x^1$ axis:
\begin{align}
\begin{pmatrix}x'^0\\x'^1\\x'^2\end{pmatrix} = \begin{pmatrix}1&0&0\\-v&1&0\\0&0&1 \end{pmatrix} \begin{pmatrix}x^0\\x^1\\x^2\end{pmatrix}.
\end{align}
Thus, the total action becomes:
\begin{align}\nonumber
S=&S^0+S^{\mathrm{int}}\nonumber\\
=&\int \mathrm{d}^3x\psi_\mathrm{L}^{\dagger} [\mathrm{i}\partial_0 - v_\mathrm{F} \sigma^1 (-\mathrm{i}\partial_1)-v_\mathrm{F} \sigma^2 (-\mathrm{i}\partial_2)-\sigma^3v_\mathrm{F}m]\psi_\mathrm{L}\nonumber\\&
+\int \mathrm{d}^3x\psi_\mathrm{R}^{\dagger} [\mathrm{i}\partial_0+v(\mathrm{i}\partial_1) - v_\mathrm{F} \sigma^1 (-\mathrm{i}\partial_1)-v_\mathrm{F} \sigma^2 (-\mathrm{i}\partial_2)-\sigma^3v_\mathrm{F}m]\psi_\mathrm{R}\nonumber\\&
+g \int \mathrm{d}^3x \int \mathrm{d}^3y [\psi^{\dagger}_\mathrm{L}(x) U(x,y) \psi_\mathrm{R}(y) +h.c.].
\end{align}
By the Fourier transformations $\psi_\mathrm{R}(x)=\int\frac{\mathrm{d}^3k}{(2\pi)^3}\psi_\mathrm{R}(k)\exp(\mathrm{i}k\cdot x)$ and $\psi_\mathrm{L}(x)=\int\frac{\mathrm{d}^3k}{(2\pi)^3}\psi_\mathrm{L}(k)\exp(\mathrm{i}k\cdot x)$, we can achieve the total action in momentum space, expressed as:
\begin{align}\nonumber
S\equiv &S[\psi_\mathrm{R},\psi_\mathrm{L}]=S^0+S^{\mathrm{int}}\nonumber\\
=&\int \frac{\mathrm{d}^3k}{(2\pi)^3}\psi_\mathrm{L}^{\dagger}(k) [k^0 + v_\mathrm{F} \sigma^1 k^1+v_\mathrm{F} \sigma^2 k^2-\sigma^3v_\mathrm{F}m]\psi_\mathrm{L}(k)\nonumber\\&
+\int \frac{\mathrm{d}^3k}{(2\pi)^3}\psi_\mathrm{R}^{\dagger}(k) [k^0-vk^1 + v_\mathrm{F} \sigma^1 k^1+v_\mathrm{F} \sigma^2 k^2-\sigma^3v_\mathrm{F}m]\psi_\mathrm{R}(k)\nonumber\\&
+g \int \frac{\mathrm{d}^3k}{(2\pi)^3} \left[\frac{\psi^{\dagger}_\mathrm{L}(k) \psi_\mathrm{R}(k)}{\sqrt{k^\mu k_\mu +\mathrm{i}\epsilon}} +h.c.\right]\nonumber\\
=&\int \frac{\mathrm{d}^3k}{(2\pi)^3}\begin{pmatrix}c_{\mathrm{L}\uparrow}^\dagger(k)&c_{\mathrm{L}\downarrow}^\dagger(k)\end{pmatrix}\begin{pmatrix}k^0-v_\mathrm{F}m&v_F(k^1-\mathrm{i}k^2)\\v_F(k^1+\mathrm{i}k^2)&k^0+v_\mathrm{F}m\end{pmatrix}\begin{pmatrix}c_{\mathrm{L}\uparrow}(k)\\c_{\mathrm{L}\downarrow}(k)\end{pmatrix}\nonumber\\
&+\int \frac{\mathrm{d}^3k}{(2\pi)^3}\begin{pmatrix}c_{\mathrm{R}\uparrow}^\dagger(k)&c_{\mathrm{R}\downarrow}^\dagger(k)\end{pmatrix}\begin{pmatrix}k^0-vk^1-v_\mathrm{F}m&v_\mathrm{F}(k^1-ik^2)\\v_\mathrm{F}(k^1+\mathrm{i}k^2)&k^0-vk^1+v_\mathrm{F}m\end{pmatrix}\begin{pmatrix}c_{\mathrm{R}\uparrow}(k)\\c_{\mathrm{R}\downarrow}(k)\end{pmatrix}\nonumber\\
&+g\int \frac{\mathrm{d}^3k}{(2\pi)^3}\left[\frac{1}{\sqrt{k^\mu k_\mu +\mathrm{i}\epsilon}}\begin{pmatrix}c_{\mathrm{L}\uparrow}^\dagger(k)&c_{\mathrm{L}\downarrow}^\dagger(k)\end{pmatrix} \begin{pmatrix}c_{\mathrm{R}\uparrow}(k)\\c_{\mathrm{R}\downarrow}(k)\end{pmatrix} +h.c.\right],
\end{align}
which is one of the key points to calculate the dissipation effect induced by relative motion in our model.

\section{Time evolution and particle production}\label{particle_production}
Next, we focus on the time evolution of the system. In general, the time evolution can be considered adiabatic, meaning that if the system is isolated ($v=0$), it would always return to the initial vacuum state after a total evolution time $T$. To describe the interaction adiabatically, we set the initial state of the system expressed in Dirac notation as the free vacuum state $| i \rangle=|0\rangle$ at time $-\frac{T}{2}$, and satisfying orthogonality $\langle 0 | 0 \rangle=1$~\cite{Weinberg101063}. Thus, the final state $|f\rangle$ can be written as:
\begin{align}
|f\rangle=| f_0 \rangle =\mathcal{U}_\mathrm{I}(\frac{T}{2},-\frac{T}{2})|0\rangle=|0\rangle \exp(\mathrm{i}\Gamma_0),
\end{align}
where $\Gamma_0$ is a real valued object and $\mathcal{U}_\mathrm{I}(\frac{T}{2},-\frac{T}{2})$ is the time evolution operator in interaction picture~\cite{Hollik101063}. The overlap between initial and final state reads:
\begin{align}
 \langle i | f_0 \rangle =\langle 0 |\mathcal{U}_\mathrm{I}(\frac{T}{2},-\frac{T}{2})|0\rangle= \langle 0 | 0 \rangle \exp(\mathrm{i}\Gamma_0) = \exp(\mathrm{i}\Gamma_0).
\end{align}
In an open system where an external force acts on the R-plate to maintain relative motion, the system’s state cannot obviously return to the initial vacuum state at the end. Instead, the final state should include all possible states, meaning that $\Gamma$ is not real-valued, resulting in an unstable vacuum.
To tell the detailed story, we define the eigenstate of the system's free Hamiltonian as $|n\rangle$, with the complete set of states:
\begin{align}
\sum_n| n \rangle\langle n |= \mathbb{I}.
\end{align}
Then, the normalization of the final state can be expressed as:
\begin{align} \nonumber
\langle f | f \rangle = &\langle 0 | \mathcal{U}^\dagger_\mathrm{I}(\frac{T}{2},-\frac{T}{2})   \mathcal{U}_\mathrm{I}(\frac{T}{2},-\frac{T}{2})|0\rangle\\
=&\sum_n\langle 0 | \mathcal{U}^\dagger_\mathrm{I}(\frac{T}{2},-\frac{T}{2})    | n \rangle\langle n |      \mathcal{U}_\mathrm{I}(\frac{T}{2},-\frac{T}{2})|0\rangle\nonumber  \\
=&\langle 0 | \mathcal{U}^\dagger_\mathrm{I}(\frac{T}{2},-\frac{T}{2})    | 0 \rangle\langle 0 |      \mathcal{U}_\mathrm{I}(\frac{T}{2},-\frac{T}{2})|0\rangle\nonumber 
+\sum_{n>0}\langle 0 | \mathcal{U}^\dagger_\mathrm{I}(\frac{T}{2},-\frac{T}{2})    | n \rangle\langle n |      \mathcal{U}_\mathrm{I}(\frac{T}{2},-\frac{T}{2})|0\rangle\nonumber\\
=&1.
\end{align}
The terms for $n>0$ refer to the probability of particle production(the bound states are omitted for simplicity),  so $\langle 0 | \mathcal{U}^\dagger_\mathrm{I}(\frac{T}{2},-\frac{T}{2}) | 0 \rangle\langle 0 |\mathcal{U}_\mathrm{I}(\frac{T}{2},-\frac{T}{2})|0\rangle$ is regarded as the transition probability from vacuum to vacuum, i.e. the vacuum persistence probability~\cite{Xu1474579}. And the corresponding amplitude $ Z=\langle 0 |      \mathcal{U}_\mathrm{I}(\frac{T}{2},-\frac{T}{2})|0\rangle=\mathrm{e}^{\mathrm{i}\Gamma}$ is called Vacuum Persistence Amplitude(VPA). Thus, the probability of particle production can be written as:
\begin{align}\nonumber
\mathcal{P}=&1- \langle 0 | \mathcal{U}^\dagger_\mathrm{I}(\frac{T}{2},-\frac{T}{2})    | 0 \rangle\langle 0 |      \mathcal{U}_\mathrm{I}(\frac{T}{2},-\frac{T}{2})|0\rangle \nonumber\\
=&1-\exp(-\mathrm{i}\Gamma^*)\exp(\mathrm{i}\Gamma) \nonumber\\
=&1-\exp(-2\mathrm{Im}\Gamma),
\label{PPP}
\end{align}
and the relative motion induces particle production in the system, which means it has on-shell excitations in both the L- and the R-plate.
Now, the final state can be expressed as:
\begin{align}
|f\rangle = |0\rangle\langle 0|\mathcal{U}_\mathrm{I}(\frac{T}{2},-\frac{T}{2})|0\rangle+  \sum_{n>0}|n\rangle\langle n|\mathcal{U}_\mathrm{I}(\frac{T}{2},-\frac{T}{2})|0\rangle,
\end{align}
and the final vacuum state can be written as:
\begin{align}
|\Omega \rangle=|0\rangle\exp(\mathrm{i}\mathrm{Re}\Gamma)\exp(-\mathrm{Im}\Gamma),
\end{align}
which is different from the initial vacuum. And here the factor $\exp(-\mathrm{Im}\Gamma)$ changes the modulus square of $|\Omega \rangle$.

In this work, we consider only the case where the final state is not far from the initial state, which can be analyzed using perturbative methods.
This implies that the terms with $n>0$ in the final state should be small enough.
In addition, the interaction within the system leads to a finite lifetime of the excitations, resulting in the system reaching a non-equilibrium steady state (NESS). Furthermore, the external force acting on the R-plate excites particles in the system, and, due to the interaction, the particles decay. These two processes occur simultaneously, leading the system to reach mechanical equilibrium.

Generally, the VPA is expressed in Heisenberg picture via asymptotic states~\cite{Weinberg101063}. So firstly we define the asymptotic vacuum states $|\mathrm{VAC,in} \rangle$ and $|\mathrm{VAC,out} \rangle$ as:
\begin{align}
&|\mathrm{VAC,in} \rangle= \mathcal{U}_{\mathrm{I}} (0,-\frac{T}{2}) |0\rangle,\\
&|\mathrm{VAC,out} \rangle= \mathcal{U}_{\mathrm{I}}^\dagger (\frac{T}{2},0) |0\rangle,
\end{align}
and then the VPA can be written as:
\begin{align}
Z=\langle\mathrm{VAC,out}|\mathrm{VAC,in}\rangle = \mathrm{e}^{\mathrm{i}\Gamma}.
\label{VPA}
\end{align}
The vacuum expectation value of an operator $\mathcal{O}_\mathrm{H}$ in Heisenberg picture can be expressed as:
\begin{align}
\langle\mathrm{VAC,in}|\mathcal{O}_\mathrm{H}|\mathrm{VAC,in}\rangle = \frac{\langle\mathrm{VAC,out}|\mathcal{O}_\mathrm{H}|\mathrm{VAC,in}\rangle}{\langle\mathrm{VAC,out}|\mathrm{VAC,in}\rangle}.
\end{align}
Here we neglect the contribution from the excited states with $n>0$ in Eq.~\eqref{PPP}, since we are only focusing on the NESS case that is near the initial vacuum state.

\section{Excitations of L-plate}\label{excitations_L}
In the previous section, the formal equation of motion-induced particle production was given. Now, we focus on the vacuum occupation number in momentum space, including quantum effects. The relative motion induces on-shell excitations in both the L-plate and the R-plate. Here, we consider only the L-plate, but the impact of the R-plate cannot be disregarded. To obtain the effective action of the L-electrons under the influence of the R-plate, we integrate out the R-electrons in the path integral formulation of the system as~\cite{Hollik101063}:
\begin{align}
\exp(\mathrm{i}S_{\mathrm{eff}}[\psi_\mathrm{L},\psi_\mathrm{L}^\dagger])=\int \mathrm{D}\psi_\mathrm{R} \mathrm{D}\psi_\mathrm{R}^\dagger \exp(\mathrm{i}S),
\end{align}
and then the effective action of L-electrons can be written as:
\begin{align}\nonumber
S_{\mathrm{eff}}[\psi_\mathrm{L},\psi_\mathrm{L}^\dagger]
=&\int\frac{\mathrm{d}^3k}{(2\pi)^3}\begin{pmatrix}c_{\mathrm{L}\uparrow}(k)\\c_{\mathrm{L}\downarrow}(k)\end{pmatrix}^\dagger\begin{pmatrix}k^0-v_\mathrm{F}m&v_\mathrm{F}(k^1-\mathrm{i}k^2)\\v_\mathrm{F}(k^1+\mathrm{i}k^2)&k^0+v_\mathrm{F}m\end{pmatrix}\begin{pmatrix}c_{\mathrm{L}\uparrow}(k)\\c_{\mathrm{L}\downarrow}(k)\end{pmatrix}\nonumber\\&+g^2\int \frac{\mathrm{d}^3k}{(2\pi)^3}\left[\frac{1}{k^\mu k_\mu +\mathrm{i}\epsilon}\begin{pmatrix}c_{\mathrm{L}\uparrow}(k)\\c_{\mathrm{L}\downarrow}(k)\end{pmatrix}^\dagger
\begin{pmatrix}k^0-vk^1-v_\mathrm{F}m&v_\mathrm{F}(k^1-\mathrm{i}k^2)\\v_\mathrm{F}(k^1+\mathrm{i}k^2)&k^0-vk^1+v_\mathrm{F}m\end{pmatrix}^{-1}
\begin{pmatrix}c_{\mathrm{L}\uparrow}(k)\\c_{\mathrm{L}\downarrow}(k)\end{pmatrix}\right].
\end{align}
To derive the vacuum occupation number of the system in momentum space perturbatively, we start with the particle number density operator, defined as:
\begin{align}
\hat{n}_\mathrm{L}(x^0,\emph{\textbf{x}}):= \psi_\mathrm{L}^\dagger(x^0,\emph{\textbf{x}}) \psi_\mathrm{L}(x^0,\emph{\textbf{x}}).
\end{align}
So the vacuum expectation value can be written as:
\begin{align}\nonumber
n_\mathrm{L}(x^0,\emph{\textbf{x}}):=&
\langle \mathrm{VAC,in} | \psi_\mathrm{L}^\dagger(x^0,\emph{\textbf{x}}) \psi_\mathrm{L}(x^0,\emph{\textbf{x}}) |\mathrm{VAC,in} \rangle\nonumber\\
=&\lim_{y\rightarrow x} \mathrm{tr}\langle \mathrm{VAC,in} | \mathcal{T}[\psi_\mathrm{L}(x^0,\emph{\textbf{x}}) \psi_\mathrm{L}^\dagger(y^0,\emph{\textbf{y}}) ] |\mathrm{VAC,in} \rangle \nonumber\\
=&\lim_{y\rightarrow x}\frac{\mathrm{tr}\langle \mathrm{VAC,out} | \mathcal{T}[\psi_\mathrm{L}(x^0,\emph{\textbf{x}}) \psi_\mathrm{L}^\dagger(y^0,\emph{\textbf{y}}) ] |\mathrm{VAC,in} \rangle}{\langle \mathrm{VAC,out} | \mathrm{VAC,in}\rangle}\nonumber\\
=&-\mathrm{i}\lim_{y\rightarrow x} \mathrm{tr} \Delta_\mathrm{L}(x,y).
\end{align}
Here `$\mathrm{tr}$' means tracing over the spin indices, and $\mathcal{T}$ is the time-ordered operator. $\Delta_\mathrm{L}(k^0,\emph{\textbf{k}})$ is the exact propagator of the L-electrons.

Since the state of the system is a non-equilibrium steady state (NESS) and homogeneous, the exact propagator in momentum space does not depend on $\frac{x+y}{2}$. This means that the Fourier transform can be directly applied, as follows:
\begin{align}
n_\mathrm{L}(x^0,\emph{\textbf{x}})
=-\mathrm{i}  \lim_{y\rightarrow x}    \int\frac{\mathrm{d}k^0}{2\pi} \int\frac{\mathrm{d}^2k}{(2\pi)^2} \mathrm{tr}[\Delta_L(k^0,\emph{\textbf{k}})] \exp(-\mathrm{i}k\cdot(x-y)).
\end{align}
After performing the $k^0$-integration, the corresponding expression for the vacuum occupation number in momentum space is given by:
\begin{align}
n_\mathrm{L}(\emph{\textbf{k}})=-\mathrm{i} \int\frac{\mathrm{d}k^0}{2\pi} \mathrm{tr}[\Delta_\mathrm{L}(k^0,\emph{\textbf{k}})].
\end{align}
We calculate the propagator $\Delta_\mathrm{L}(k^0,\emph{\textbf{k}})$ perturbatively up to order $o(g^2)$, and the result is:
\begin{align}
\Delta_\mathrm{L}(k^0,\emph{\textbf{k}})=G_\mathrm{L}(k^0,\emph{\textbf{k}})+\frac{g^2}{k^\mu k_\mu +\mathrm{i}\epsilon} G_\mathrm{L}(k^0,\emph{\textbf{k}}) G_\mathrm{R}(k^0-vk^1,\emph{\textbf{k}})  G_\mathrm{L}(k^0,\emph{\textbf{k}}).
\label{propagator}
\end{align}
The first term in Eq.~\eqref{propagator} comes from the free L-electrons, specifically a loop propagator, corresponding to the contribution of vacuum energy. In the canonical approach, this loop propagator can be removed by the normal ordering of the field operators, which means the vacuum occupation number of L-electrons is zero in the free theory. The second term in Eq.~\eqref{propagator} refers to the interaction with the R-electrons. In our study, the L-electrons evolve under an effective potential due to the integration of the R-electrons, so they can be considered as an OQS. Therefore, the vacuum occupation number of L-electrons is modified by the effective potential, meaning it is nonzero due to the interaction.

Through path integral formulas, the free electron propagators $G_{\mathrm{R/L}}(x)$ can be defined as~\cite{Hollik101063}:
\begin{align}
G_{\mathrm{R/L}}(x)=\frac {\int \mathrm{D}\psi_{\mathrm{R/L}} \mathrm{D}\psi_{\mathrm{R/L}}^{\dagger} \exp[\mathrm{i}S^0] \psi_{\mathrm{R/L}}\psi_{\mathrm{R/L}}^{\dagger}}   {\int \mathrm{D}\psi_{\mathrm{R/L}} \mathrm{D}\psi_{\mathrm{R/L}}^{\dagger} \exp[\mathrm{i}S^0]}.
\end{align}
The explicit expressions of L- and R- plates in momentum space can be written as:
\begin{align}
G_\mathrm{L}(k)= \frac{1}{(k^0)^2 - v_\mathrm{F}^2(k^1)^2 - v_\mathrm{F}^2(k^2)^2-v_\mathrm{F}^2m^2+\mathrm{i}\epsilon} \begin{pmatrix}k^0+v_\mathrm{F}m&-v_\mathrm{F}(k^1-\mathrm{i}k^2)\\-v_\mathrm{F}(k^1+\mathrm{i}k^2)&k^0-v_\mathrm{F}m\end{pmatrix},
\end{align}
and
\begin{align}\nonumber
\tilde{G}_\mathrm{R}(k^0,k^1,k^2)=&G_\mathrm{R}(k^0-vk^1,k^1,k^2)\nonumber\\
=&\frac{1}{(k^0-vk^1)^2 - v_\mathrm{F}^2(k^1)^2 - v_\mathrm{F}^2(k^2)^2-v_\mathrm{F}^2m^2+\mathrm{i}\epsilon} \begin{pmatrix}k^0-vk^1+v_\mathrm{F}m&-v_\mathrm{F}(k^1-\mathrm{i}k^2)\\-v_\mathrm{F}(k^1+\mathrm{i}k^2)&k^0-vk^1-v_\mathrm{F}m\end{pmatrix}.
\end{align}

The relative motion causes the speed to appear in the propagator, which leads to a change in the modified vacuum occupation number. To focus on the effects of relative motion, we redefine the vacuum occupation number as:
\begin{align}\nonumber
n_\mathrm{L}(\emph{\textbf{k}}):=&-\mathrm{i} \int\frac{\mathrm{d}k^0}{2\pi} \mathrm{tr}[\Delta_\mathrm{L}(k^0,\emph{\textbf{k}})-G_\mathrm{L}(k^0,\emph{\textbf{k}})]\nonumber\\
=&-\mathrm{i} \int\frac{\mathrm{d}k^0}{2\pi}\frac{g^2}{k^\mu k_\mu +\mathrm{i}\epsilon}  \mathrm{tr}[G_\mathrm{L}(k^0,\emph{\textbf{k}}) G_\mathrm{R}(k^0-vk^1,\emph{\textbf{k}})  G_\mathrm{L}(k^0,\emph{\textbf{k}})]\nonumber\\
=&2\mathrm{i}g^2\int\frac{\mathrm{d}k^0}{2\pi}\frac{1}{(k^\mu k_\mu +\mathrm{i}\epsilon)[(k^0)^2-v_\mathrm{F}^2(\emph{\textbf{k}}^2+m^2)+\mathrm{i}\epsilon]^2[(k^0-vk^1)^2-v_\mathrm{F}^2(\emph{\textbf{k}}^2+m^2)+\mathrm{i}\epsilon]}.
\end{align}
The integrand has three simple poles in the $k^0$-plane. Since in this study only the effects induced by the relative motion are considered, we focus on the simple poles that depend on the relative velocity $\emph{\textbf{v}}=\begin{pmatrix}v, 0\end{pmatrix}^\mathrm{T}$, i.e., $\Lambda_{\pm}=vk^1 \pm v_\mathrm{F} \sqrt{(k^1)^2+(k^2)^2+m^2-\mathrm{i}\epsilon} \approx vk^1 \pm v_\mathrm{F} \sqrt{(k^1)^2+(k^2)^2+m^2} \mp \mathrm{i}\epsilon$. 
The reasonable contours can be built to evaluate the integral over $k^0$ using the residue theorem as~\cite{Hollik101063}:
\begin{align}
n_\mathrm{L}(\emph{\textbf{k}})= \frac{g^2\pi(vk^1+2\omega_k)}{[(vk^1+\omega_k)^2-\emph{\textbf{k}}^2](vk^1)} + \frac{g^2\pi(vk^1-2\omega_k)}{[(vk^1-\omega_k)^2-\emph{\textbf{k}}^2](vk^1)},
\end{align}
where $\omega_k:=v_F\sqrt{\emph{\textbf{k}}^2+m^2}$ is the dispersion of free electrons. 

\begin{figure}[htbp]
\centering
\subfigure[3D-plot, $v = 0$]{
\label{Fig.0.1}
\includegraphics[height=0.4\textwidth]{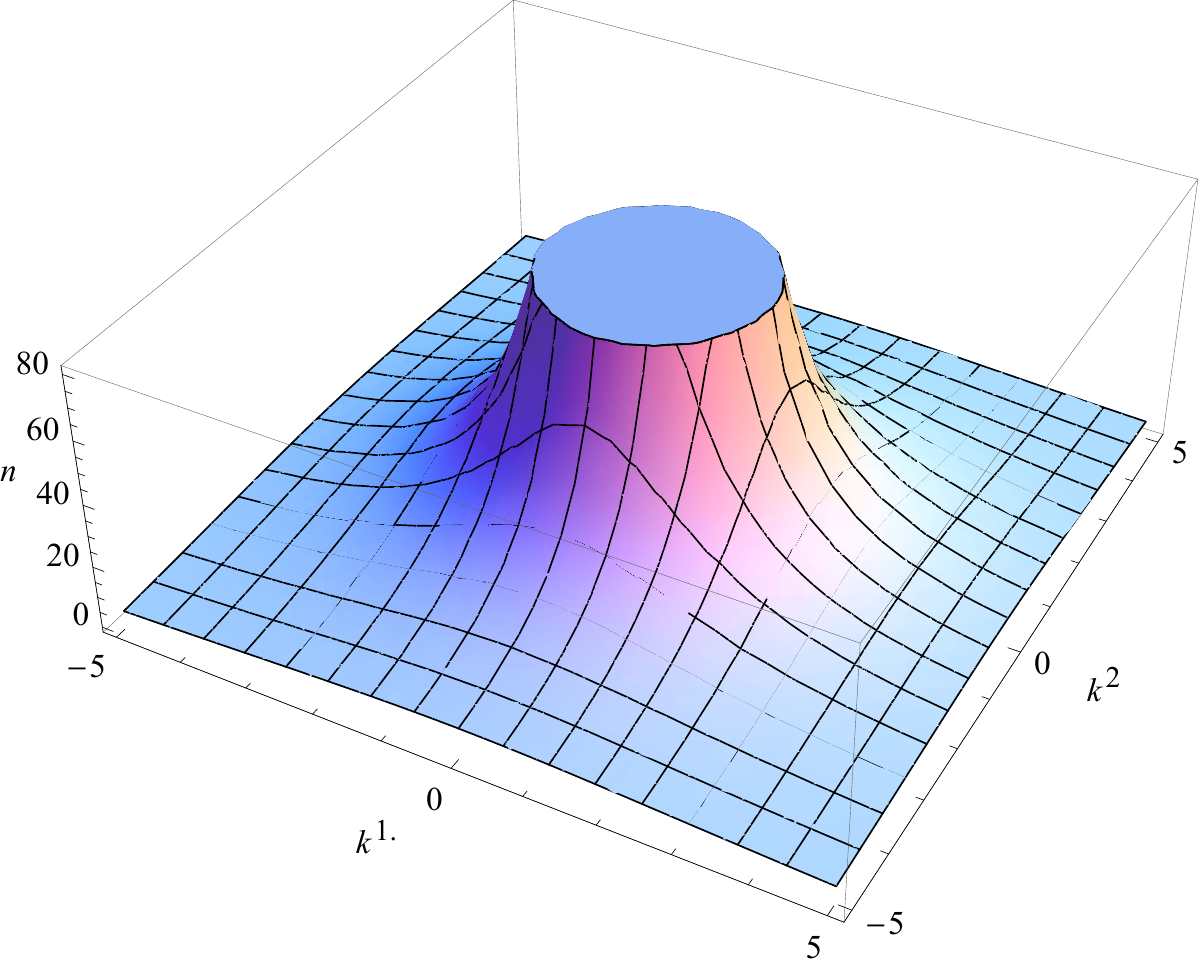}}
\subfigure[Contour-plot, $v = 0$]{
\label{Fig.0.2}
\includegraphics[height=0.4\textwidth]{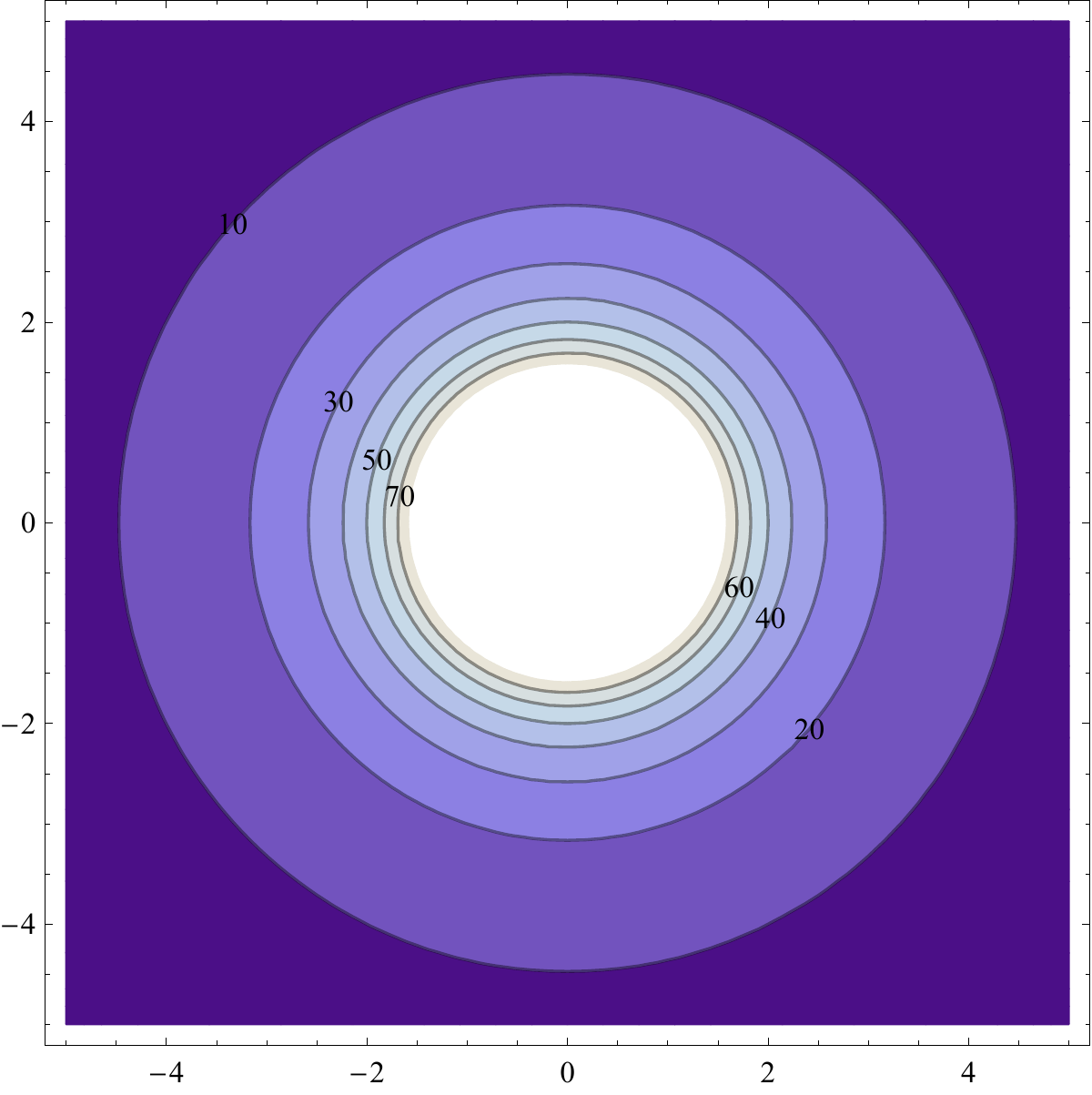}}
\subfigure[3D-plot, $v = 7$]{
\label{Fig.0.3}
\includegraphics[height=0.4\textwidth]{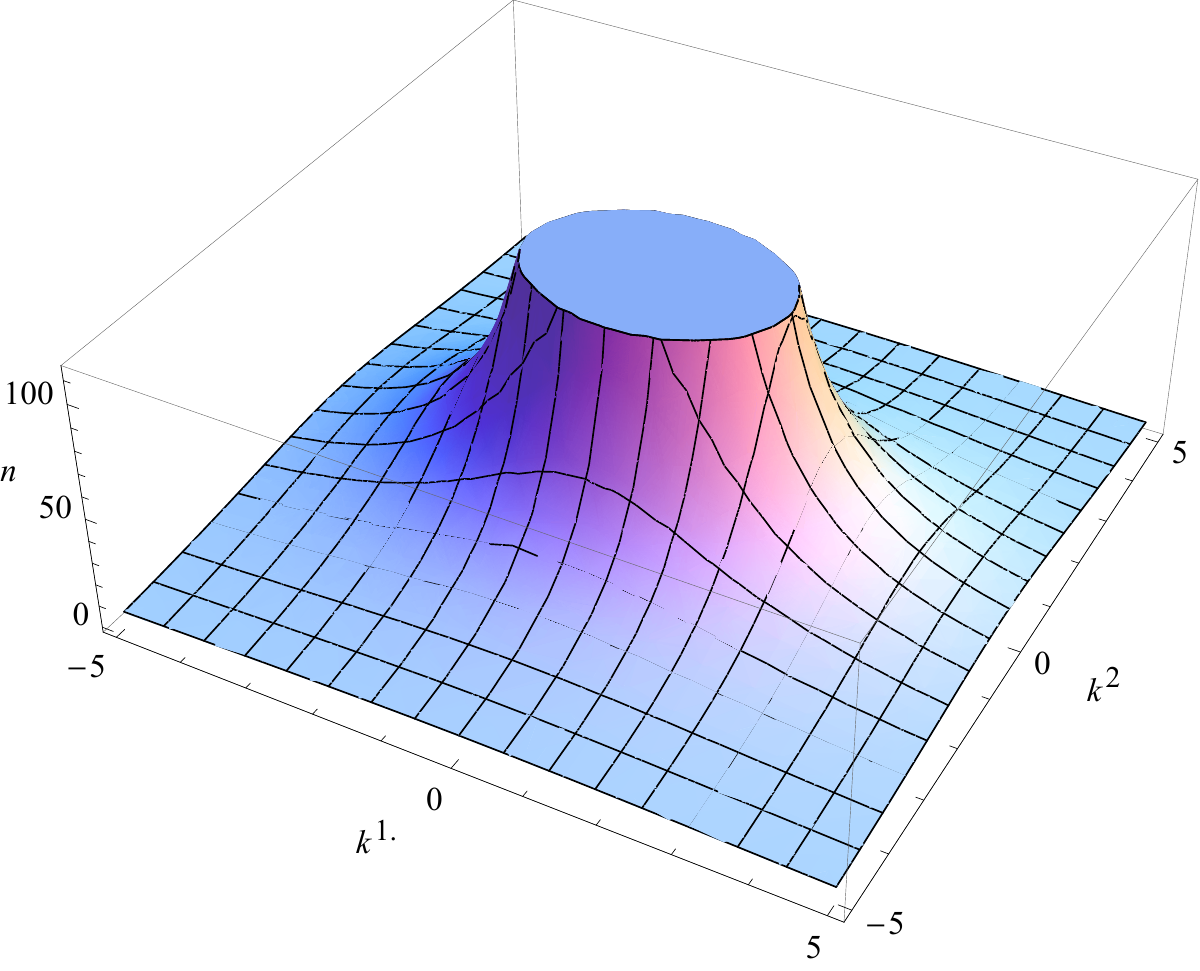}}
\subfigure[Contour-plot, $v = 7$]{
\label{Fig.0.4}
\includegraphics[height=0.4\textwidth]{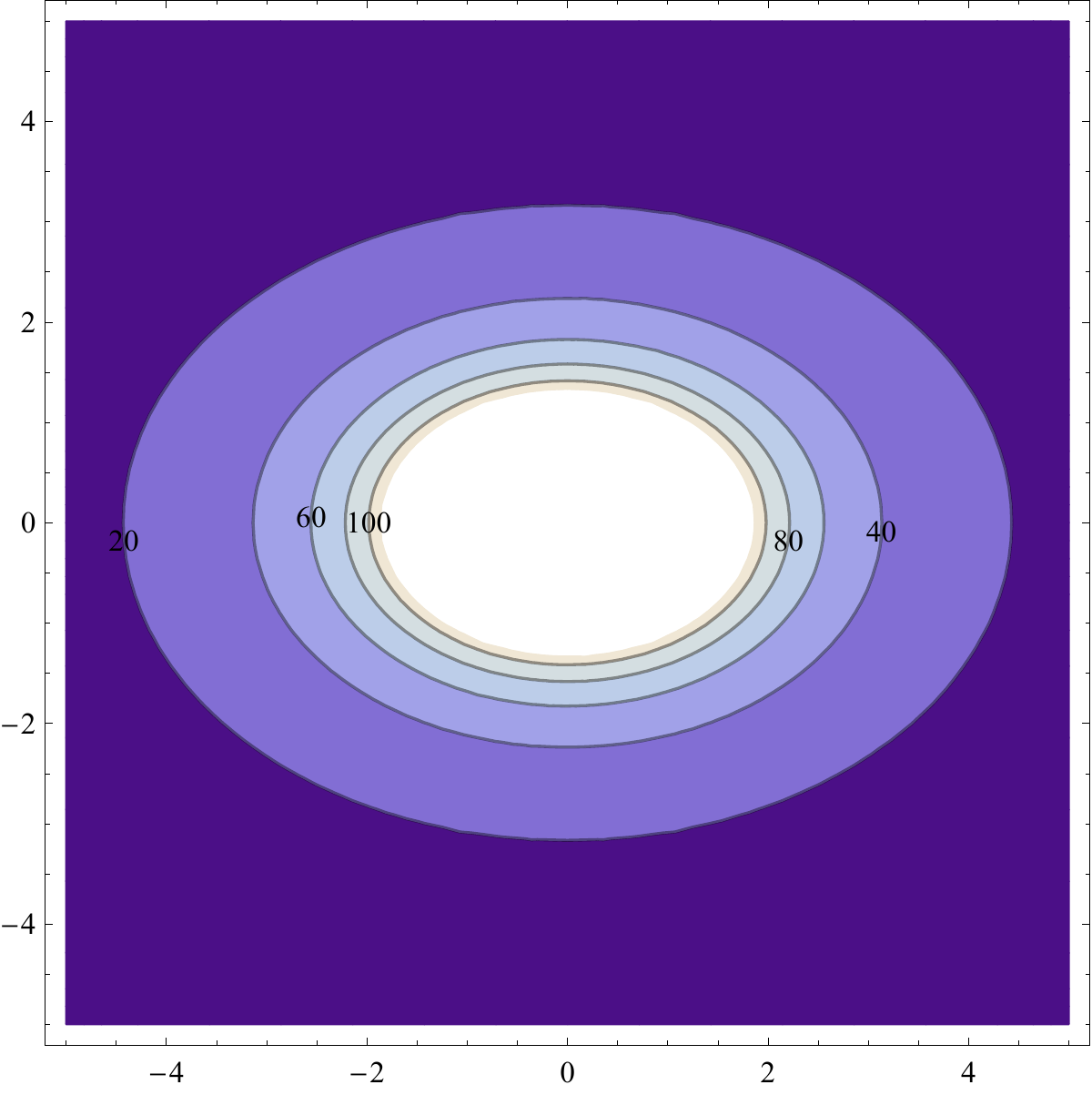}}
\caption{3D and Contour-plot of the vacuum occupation number, for the case $v_\mathrm{F}=0.001$, $m=10$, in the order of $g^2$. Here {\bf (a)}\, and {\bf (b)} for $v=0$; {\bf (c)} and {\bf (d)} for $v=7$. }
\label{numerical_contours}
\end{figure}

The numerical results of the vacuum occupation number for $v_\mathrm{F}=0.001$ and $m=10$ in the cases of $v=0$ and $v=0.7$ re shown in Fig.~\ref{numerical_contours} as examples. Comparing subfigures in the top row ($v=0$) with those in the bottom row ($v=7$),  it is evident that as $v$ increases, the occupation number is stretched along the $k_1$-direction, corresponding to the variation of $v$ along the $x^1$-axis. This result is expected because the relative motion is along the $x^1$-axis, meaning an external force acts on the R-plate in that direction, which causes the momentum of the L-electrons to vary accordingly. As a function of momentum, $n_\mathrm{L}(\emph{\textbf{k}})$ is isotropic for $v=0$, but becomes anisotropic for nonzero $v$.
Moreover, $n_\mathrm{L}(\emph{\textbf{k}})$ is time-independent, indicating that our approximation applies to a NESS regime where the system remains close to the initial vacuum state. The nonzero vacuum occupation number in momentum space shows that the interaction modifies the free vacuum, and the relative motion induces on-shell excitations from the vacuum.

\section{Quantum dissipation}\label{dissipation}
In Section~\ref{particle_production}, we discussed how relative motion induces on-shell excitations. This process is analogous to the Schwinger effect, as particle production is a dissipative process where the energy of the excitations originates from the external force~\cite{PhysRevD.73.065020}. The energy transfer within the system implies the existence of dissipative forces between the two plates. Consequently, in this section, we investigate the motion-induced dissipation effects and the corresponding dissipative forces.

\subsection{The quantum action}
From Eq.~\eqref{VPA}, the VPA can be expressed via path integral as~\cite{OLESEN1978327}:
\begin{align}
Z=\mathrm{e}^{\mathrm{i}\Gamma}=\int \mathrm{D}\psi_\mathrm{R} \mathrm{D}\psi_\mathrm{R}^{\dagger} \int \mathrm{D}\psi_\mathrm{L} \mathrm{D}\psi_\mathrm{L}^{\dagger}  \exp(\mathrm{i}S).
\end{align}
After performing the integration over $\psi_\mathrm{R}, \psi_\mathrm{R}^{\dagger}$, $\psi_\mathrm{L} and \psi_\mathrm{L}^{\dagger}$, it becomes:
\begin{align}
\exp(\mathrm{i}\Gamma)=[\det (\mathrm{i}G_\mathrm{L}^{-1}+\mathrm{i}g^2 U^2 \tilde{G}_\mathrm{R})],
\end{align}
where $\Gamma$ is the quantum action that contains all quantum information of the system. Thus, the quantum action can be written as:
\begin{align}\nonumber
\mathrm{i}\Gamma&=\ln[\det (\mathrm{i}G_\mathrm{L}^{-1}+g^2U^2\tilde{G}_\mathrm{R})]\nonumber\\&
=\mathrm{Tr}[\ln (\mathrm{i}G_\mathrm{L}^{-1}+\mathrm{i}g^2U^2\tilde{G}_\mathrm{R})]\nonumber\\&
=\mathrm{Tr}[\ln (\mathrm{i}G_\mathrm{L}^{-1})(1+g^2 U^2 G_\mathrm{L} \tilde{G}_\mathrm{R})]\nonumber\\&
=\mathrm{Tr}[\ln (\mathrm{i}G_\mathrm{L}^{-1})] + \mathrm{Tr}[\ln (1+g^2 U^2 G_\mathrm{L} \tilde{G}_\mathrm{R})].
\end{align}
Since it is very complicated to calculate $\Gamma$ directly, a more general method is to expand $\Gamma$ in powers of $g$. To the order of $o(g^2)$, the expression for $\Gamma$ is:
\begin{align}
\mathrm{i}\Gamma=\mathrm{Tr}[\ln (\mathrm{i}G_\mathrm{L}^{-1})] + \mathrm{Tr}[\ln (1+g^2 U^2 G_\mathrm{L} \tilde{G}_\mathrm{R})] \approx \mathrm{Tr}[\ln (\mathrm{i}G_\mathrm{L}^{-1})] +\mathrm{Tr}(g^2 U^2 G_\mathrm{L} \tilde{G}_\mathrm{R}).
\end{align}
Here $\mathrm{Tr}[\ln (\mathrm{i}G_\mathrm{L}^{-1})]$ is an irrelevant constant that presents the vacuum energy of the Dirac field $\{\psi_\mathrm{L}(x), \psi_\mathrm{L}^{\dagger}(x)\}$ in the absence of any interaction. 
Therefore, this irrelevant constant can be dropped, and then $\Gamma$ can be expressed as:
\begin{align}\nonumber
\Gamma=&\mathrm{i}g^2\mathrm{Tr}(U^2 G_\mathrm{L} \tilde{G}_\mathrm{R})\nonumber\\
=&\mathrm{i}g^2 VT \int \frac{\mathrm{d}^3k}{(2\pi)^3}U^2(k^0,k^1,k^2) G_\mathrm{L}(k^0,k^1,k^2) \tilde{G}_\mathrm{R}(k^0,k^1,k^2)\nonumber\\
=&\mathrm{i}g^2 VT \int \frac{\mathrm{d}^3k}{(2\pi)^3}U^2(k^0,k^1,k^2) G_\mathrm{L}(k^0,k^1,k^2) G_\mathrm{R}(k^0-vk^1,k^1,k^2)\nonumber\\
=&\mathrm{i}g^2 VT \int \frac{\mathrm{d}^3k}{(2\pi)^3} \frac{1}{(k^0)^2-(k^1)^2-(k^2)^2+\mathrm{i}\epsilon}\nonumber\\
&\times\frac{1}{(k^0)^2 - v_\mathrm{F}^2(k^1)^2 - v_\mathrm{F}^2(k^2)^2-v_\mathrm{F}^2m^2+\mathrm{i}\epsilon}\nonumber\\
&\times\frac{1}{(k^0-vk^1)^2 - v_\mathrm{F}^2(k^1)^2 - v_\mathrm{F}^2(k^2)^2-v_\mathrm{F}^2m^2+\mathrm{i}\epsilon}\nonumber\\
&\times \mathrm{tr}\left[\begin{pmatrix}k^0+v_\mathrm{F}m&-v_\mathrm{F}(k^1-\mathrm{i}k^2)\\-v_\mathrm{F}(k^1+\mathrm{i}k^2)&k^0-v_\mathrm{F}m\end{pmatrix}\begin{pmatrix}k^0-vk^1+v_\mathrm{F}m&-v_\mathrm{F}(k^1-\mathrm{i}k^2)\\-v_\mathrm{F}(k^1+\mathrm{i}k^2)&k^0-vk^1-v_\mathrm{F}m\end{pmatrix}\right].
\end{align}

\subsection{The imaginary part of the quantum action and the dissipative force}
In this section, we derive the expression for the imaginary part of the quantum action and the dissipative force between the two plates.
In our model, the R-plate is subject to an external force, rendering it an open system. Consequently, the vacuum state of the system becomes unstable, leading to particle excitations within the plates. Therefore, dissipation is associated with the imaginary part of the quantum action, $\Gamma$.
The quantum transition amplitude has a physical interpretation: its square represents the probability of the corresponding transition process. In this study, the probability of the vacuum-to-vacuum transition process is expressed as $|Z|^2$. For small $\mathrm{Im}\Gamma$, this probability can be approximated perturbatively as:
\begin{align}
|Z|^2=|\exp(\mathrm{i}\Gamma)|^2=\exp(-2\mathrm{Im}\Gamma)\thickapprox 1-2\mathrm{Im}\Gamma,
\end{align}
and the probability of particle production becomes:
\begin{align}
\mathcal{P}=1-|Z|^2=2\mathrm{Im}\Gamma.
\end{align}

To obtain the imaginary part of the quantum action, we first perform the integral over $k^0$, then the quantum action can be expressed as:
\begin{align}
\Gamma=\mathrm{i}g^2 VT \int \frac{\mathrm{d}^3k}{(2\pi)^3} f(k^0),
\end{align}
where
\begin{align}\nonumber
f(k^0)=&\frac{1}{(k^0)^2-(k^1)^2-(k^2)^2+\mathrm{i}\epsilon}\nonumber\\
&\times\frac{1}{(k^0)^2 - v_\mathrm{F}^2(k^1)^2 - v_\mathrm{F}^2(k^2)^2-v_\mathrm{F}^2m^2+\mathrm{i}\epsilon}\nonumber\\
&\times\frac{1}{(k^0-vk^1)^2 - v_\mathrm{F}^2(k^1)^2 - v_\mathrm{F}^2(k^2)^2-v_\mathrm{F}^2m^2+\mathrm{i}\epsilon}\nonumber\\
&\times \mathrm{tr}\left[\begin{pmatrix}k^0+v_\mathrm{F}m&-v_\mathrm{F}(k^1-\mathrm{i}k^2)\\-v_\mathrm{F}(k^1+\mathrm{i}k^2)&k^0-v_\mathrm{F}m\end{pmatrix}\begin{pmatrix}k^0-vk^1+v_\mathrm{F}m&-v_\mathrm{F}(k^1-\mathrm{i}k^2)\\-v_\mathrm{F}(k^1+\mathrm{i}k^2)&k^0-vk^1-v_\mathrm{F}m\end{pmatrix}\right].
\label{f_k_0}
\end{align}
In this case, $f(k^0)$ has six single poles in the $k^0$ complex plane, corresponding to on-shell particles, i.e., real particles. In this work, we study motion-dependent particle production, which requires selecting the poles based on the motion speed $v$, corresponding to the particles excited by the relative motion. These $v$-dependent poles of the function $f(k^0)$ can be expressed as:
\begin{align}
\Lambda_{\pm}=vk^1 \pm v_\mathrm{F} \sqrt{(k^1)^2+(k^2)^2+m^2-\mathrm{i}\epsilon} \approx vk^1 \pm v_\mathrm{F} \sqrt{(k^1)^2+(k^2)^2+m^2} \mp \mathrm{i}\epsilon,
\end{align}
and the corresponding residues are given by:
\begin{align}\nonumber
\mathrm{Res}[f(k^0), \Lambda_{\pm}] =&\frac{1}{vk^1\pm 2v_\mathrm{F}\sqrt{(k^1)^2+(k^2)^2+m^2}\mp \mathrm{i}\epsilon}\nonumber\\
&\times\frac{\pm 1}{2v_\mathrm{F}\sqrt{(k^1)^2+(k^2)^2+m^2}\mp \mathrm{i}\epsilon} \times \frac{vk^1-2v_\mathrm{F}\sqrt{(k^1)^2+(k^2)^2+m^2}}{vk^1}\nonumber\\
&\times \frac{1}{(vk^1)[vk^1\pm 2v_\mathrm{F}\sqrt{(k^1)^2+(k^2)^2+m^2}]+ v_\mathrm{F}^2[(k^1)^2+(k^2)^2+m^2]-(k^1)^2-(k^2)^2\mp \mathrm{i}\epsilon}.
\end{align}
Considering a contour formed by a semicircle with a very large radius in the bottom half of the complex plane and the real axis, we can complete the integral over $k^0$ using the residue theorem~\cite{PhysRevD.84.025011}. Therefore, the quantum action can be written as:
\begin{align}
\Gamma = g^2VT \int \frac{\mathrm{d}k^2}{2\pi} \int \frac{\mathrm{d}k^1}{2\pi} \mathrm{Res}[f(k^0), \Lambda_{-}].
\end{align}
Both the contribution from the off-shell electrons (i.e., the vacuum fluctuation) and the on-shell electrons have been contained, so the quantum action can be written as:
\begin{align}\nonumber
\Gamma = &g^2VT \int \frac{\mathrm{d}k^2}{2\pi} \int \frac{\mathrm{d}k^1}{2\pi} \frac{1}{vk^1- 2v_\mathrm{F}\sqrt{(k^1)^2+(k^2)^2+m^2}+ \mathrm{i}\epsilon}\nonumber\\
&\times\frac{-1}{2v_\mathrm{F}\sqrt{(k^1)^2+(k^2)^2+m^2}+ \mathrm{i}\epsilon} \times \frac{vk^1-2v_\mathrm{F}\sqrt{(k^1)^2+(k^2)^2+m^2}}{vk^1}\nonumber\\
&\times \frac{1}{(vk^1)[vk^1+ 2v_\mathrm{F}\sqrt{(k^1)^2+(k^2)^2+m^2}]+ v_\mathrm{F}^2[(k^1)^2+(k^2)^2+m^2]-(k^1)^2-(k^2)^2+ \mathrm{i}\epsilon}.
\end{align}

\begin{figure}
  \centering
  \includegraphics[width=3in]{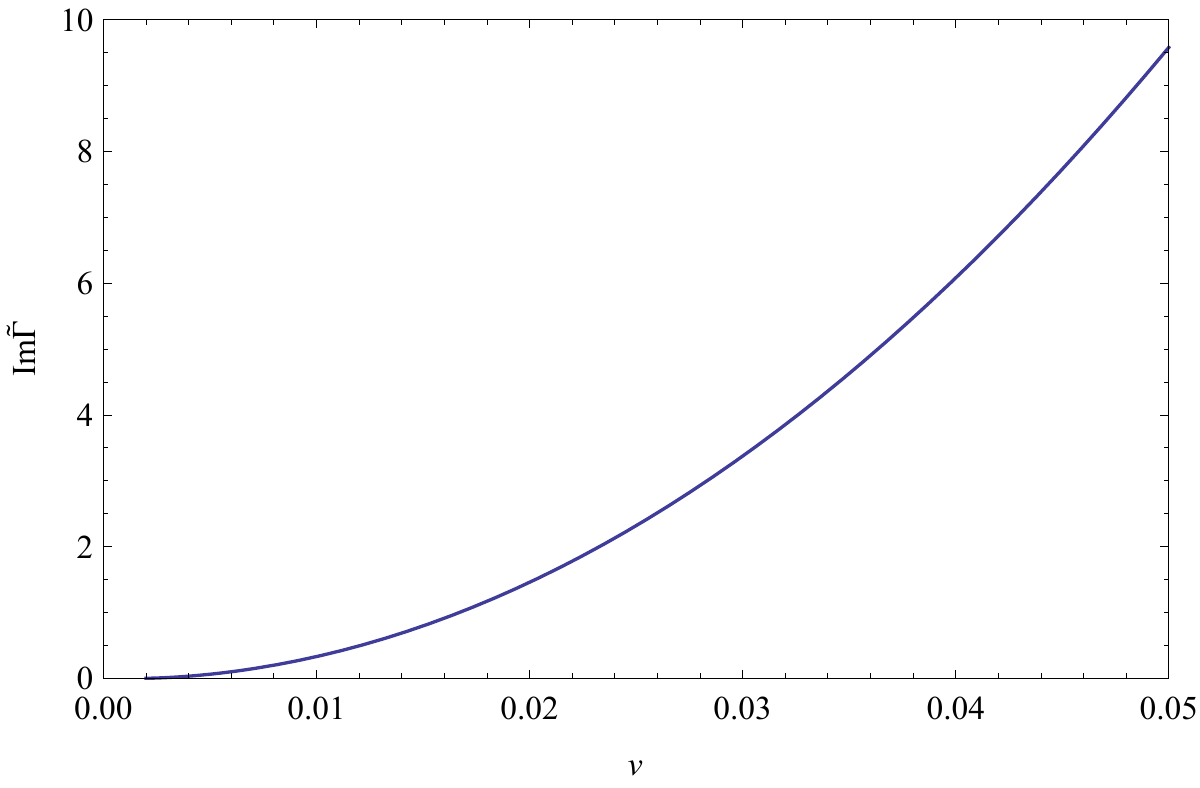}\\
  \caption{The imaginary part of quantum action as a function of $v$, for the case $v_\mathrm{F}=0.001$, $m=10$, in units of $g^2TV$.}\label{Im_part}
\end{figure}

The imaginary part of $\Gamma$ is:
\begin{align}\nonumber
\mathrm{Im}\Gamma =& g^2VT \int \frac{\mathrm{d}k^2}{2\pi} \int \frac{\mathrm{d}k^1}{2\pi} \mathrm{Im} \left(\frac{1}{vk^1- 2v_\mathrm{F}\sqrt{(k^1)^2+(k^2)^2+m^2}+ \mathrm{i}\epsilon}\right)\nonumber\\
&\times\frac{-1}{2v_\mathrm{F}\sqrt{(k^1)^2+(k^2)^2+m^2}+ \mathrm{i}\epsilon} \times \frac{vk^1-2v_\mathrm{F}\sqrt{(k^1)^2+(k^2)^2+m^2}}{vk^1}\nonumber\\
&\times \frac{1}{(vk^1)[vk^1+ 2v_\mathrm{F}\sqrt{(k^1)^2+(k^2)^2+m^2}]+ v_\mathrm{F}^2[(k^1)^2+(k^2)^2+m^2]-(k^1)^2-(k^2)^2+ \mathrm{i}\epsilon}.
\end{align}
For any real function $\frac{1}{a}$, it has:
\begin{align}
\lim_{\epsilon\rightarrow 0}\frac{1}{a\pm \mathrm{i}\epsilon} = \textbf{p.v.}(\frac{1}{a}) \mp \mathrm{i}\pi\delta(a),
\end{align}
where $\textbf{p.v.}$ means the principal value. This indictates that  $\mathrm{Im}\left(\frac{1}{a+\mathrm{i}\epsilon}\right) = \pi \delta(a)$, and then we can write:
\begin{align}\nonumber
\mathrm{Im}\Gamma = &g^2VT \int \frac{\mathrm{d}k^2}{2\pi} \int \frac{\mathrm{d}k^1}{2\pi} \delta[vk^1 - 2v_\mathrm{F}\sqrt{(k^1)^2+(k^2)^2+m^2}]\nonumber\\
&\times\frac{-1}{2v_\mathrm{F}\sqrt{(k^1)^2+(k^2)^2+m^2}+ \mathrm{i}\epsilon} \times \frac{vk^1-2v_\mathrm{F}\sqrt{(k^1)^2+(k^2)^2+m^2}}{vk^1}\nonumber\\
&\times \frac{1}{(vk^1)[vk^1+ 2v_\mathrm{F}\sqrt{(k^1)^2+(k^2)^2+m^2}]+ v_\mathrm{F}^2[(k^1)^2+(k^2)^2+m^2]-(k^1)^2-(k^2)^2+ \mathrm{i}\epsilon}.
\label{im_part_1}
\end{align}
The imaginary part of $\Gamma$ represents dissipation as well as the production of electrons from the initial vacuum. Relative motion causes energy transfer between electrons and introduces imaginary components in theoretical calculations. The energy of a single electron transforms as $\omega'_k = \omega_k + vk^1$ under the Galilean boost of the coordinate system, where $vk^1$ can be interpreted as the energy supplied to the system by the external force. From Eq.~\eqref{im_part_1}, it follows that both plates have an excited electron with energy $vk^1$ due to the transformation.
By the properties of the Dirac $\delta$-function, we know that:
\begin{align}
\delta[vk^1 - 2v_\mathrm{F}\sqrt{(k^1)^2+(k^2)^2+m^2}]=\frac{v}{v^2-4v_\mathrm{F}^2}[\delta(k^1-2v_\mathrm{F}\sqrt{\frac{(k^2)^2+m^2}{v^2-4v_\mathrm{F}^2}})+\delta(k^1+2v_\mathrm{F}\sqrt{\frac{(k^2)^2+m^2}{v^2-4v_\mathrm{F}^2}})].
\end{align}
Thus, the imaginary part of $\Gamma$, after completing the integral over $k^1$, can be expressed as:
\begin{align}
\mathrm{Im}\Gamma = \frac{g^2v_\mathrm{F}VT}{4} \sqrt{v^2-4v_\mathrm{F}^2} \int \frac{\mathrm{d}k^2}{2\pi} \frac{1}{\sqrt{v_\mathrm{F}^2(k^2)^2+v_\mathrm{F}^2m^2}}\frac{1}{(1-9v_\mathrm{F}^2)v^2(k^2)^2-(9v^2-4)v_\mathrm{F}^2m^2}.
\label{Im_Gamma}
\end{align}
In Fig.~\ref{Im_part}, we give the numerical result of Eq.~\eqref{Im_Gamma} for $v_\mathrm{F}=0.001$ and $m=10$, which shows a threshold of the relative velocity $v=2v_\mathrm{F}=0.002$. 
This indicates that dissipation occurs for $v>2v_\mathrm{F}$. 

\begin{figure}
  \centering
  \includegraphics[width=3in]{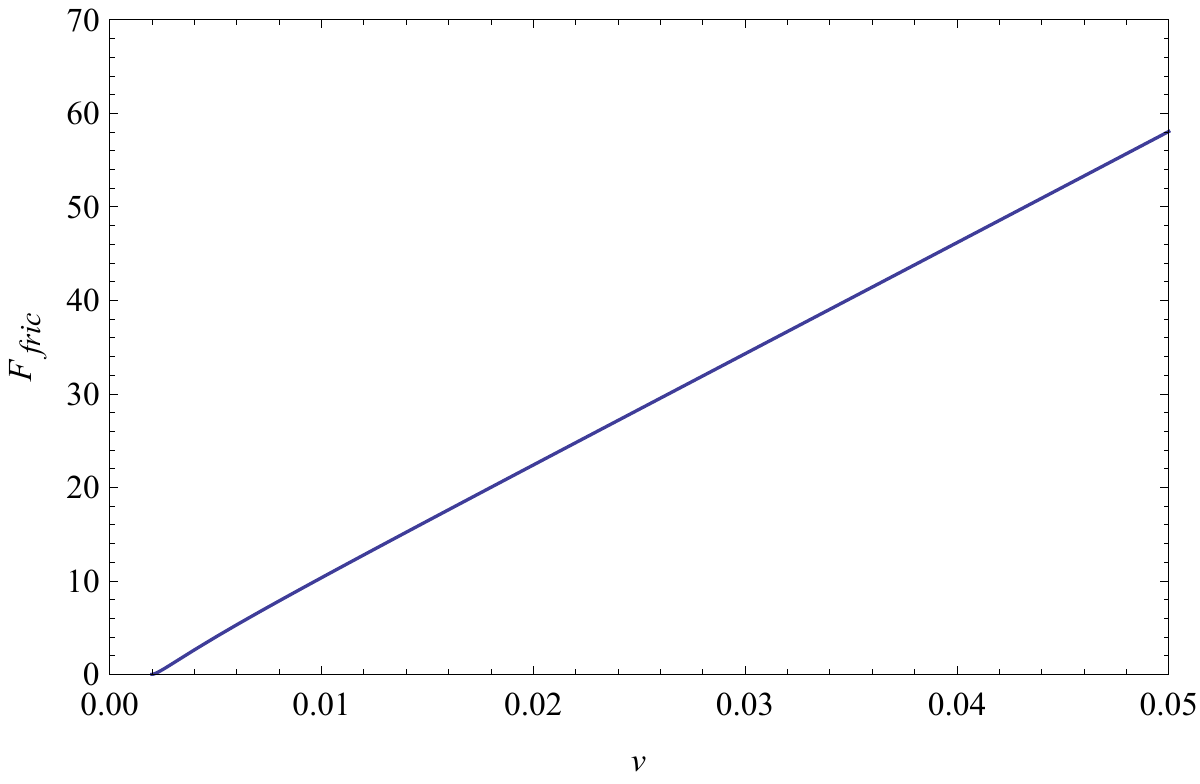}\\
  \caption{The dissipative force as a function of $v$, for the case $v_\mathrm{F}=0.001$, $m=10$, in units of $g^2$.}
  \label{Diss}
\end{figure}

The dissipation occurring in the system refers to an energy transfer process, since it is an external force acting on the R-plate. As the relative motion progresses, energy is continuously pumped into the system. This energy excites electrons at the interface between both plates. Thus, the equation of energy transfer can be written as:
\begin{align}
\frac{E_{\mathrm{pump}}}{VT}=\frac{E_{\mathrm{diss}}}{VT},
\end{align}
which indicates the power of energy pumped into the system ($E_{\mathrm{pump}}$) equals the dissipation power ($E_{\mathrm{diss}}$). To maintain the relative motion between the two plates, a continuous external force must be applied to the R-plate to counteract the frictional force between the plates. Only in this way can the external source resist friction and maintain uniform relative motion. The dissipation energy from the excitation of electrons can be expressed as:
\begin{align}
E_{\mathrm{diss}}=g^2VT \mathrm{Im}\int \frac{\mathrm{d}^3k}{(2\pi)^3} \mathrm{i}f(k^0) |k^0|,
\end{align}
and the dissipation power can be written as:
\begin{align}
\frac{E_{\mathrm{diss}}}{VT}=F_{\mathrm{diss}} v.
\end{align}
Thus, we can achieve the dissipative force as:
\begin{align}
F_{\mathrm{diss}} =\frac{1}{v} g^2  \mathrm{Im} \int \frac{\mathrm{d}^3k}{(2\pi)^3} \mathrm{i}f(k^0) |k^0|.
\end{align}
The explicit expression of the dissipative force can be obtained from Eq.~\eqref{f_k_0} as:
\begin{align}
F_{\mathrm{diss}} = \frac{3g^2v_\mathrm{F}}{4v} \sqrt{v^2-4v_\mathrm{F}^2} \int \frac{\mathrm{d}k^2}{2\pi} \frac{1}{\sqrt{v_\mathrm{F}^2(k^2)^2+v_\mathrm{F}^2m^2}}\frac{1}{(1-9v_\mathrm{F}^2)v^2(k^2)^2-(9v^2-4)v_\mathrm{F}^2m^2}.
\end{align}
Fig.~\ref{Diss} shows the numerical result of $F_{\mathrm{diss}}$ for $v_\mathrm{F}=0.001$ and $m=10$. We can see that it is a threshold of the relative velocity $v=2v_\mathrm{F}=0.002$, like the $v$-dependence of $\mathrm{Im}\Gamma$ in Fig.~\ref{Im_part}. The difference is, the in Fig.~\ref{Diss} the curve of $F_{\mathrm{diss}}$ shows a linear behavior, and the slope of the $v-F_{\mathrm{diss}}$ curve gives a linear friction coefficient at zero temperature.

To deeply undersand the existence of thresholds for $\mathrm{Im}\Gamma$ and $F_{\mathrm{diss}}$, we can focus on the Dirac $\delta$-function $\delta(vk^1-2v_\mathrm{F}\sqrt{(k^1)^2+(k^2)^2+m^2})$, the resonance to the external force acting on the Dirac fields living in the plates.
This resonance implies production of the on-shell electrons, and the most easily excited electrons are the massless ones, i.e., $2v_\mathrm{F}\sqrt{(k^1)^2+(k^2)^2}$.

\section{Conclusion}\label{conclusion}
In this paper, we studied the excitation and quantum dissipation induced by the internal relative motion of two parallel metallic plates. The degrees of freedom (DOFs) of the electrons in both plates were modeled using the 1+2 dimensional Dirac field, and a nonlocal potential was chosen to describe the interaction between L-electrons and R-electrons. The internal relative motion was introduced via a Galilean boost.

Specifically, by assuming the L-plate is stationary while the R-plate moves along the interface, we derived the effective action of the L-electrons by integrating out the R-electrons and then calculated the vacuum occupation number $n_\mathrm{L}(\emph{\textbf{k}})$ in momentum space perturbatively. According to the numerical results, $n_\mathrm{L}(\emph{\textbf{k}})$ is isotropic with respect to momentum for $v=0$ and becomes anisotropic for nonzero $v$, as expected. The perturbative approximation applied here operates in a nonequilibrium steady-state (NESS) regime, where the system remains close to the initial vacuum state.

Subsequently, we derived the expression for the imaginary part of the quantum action $\mathrm{Im}\Gamma$ due to the relative motion and calculated the dissipative force $F_{\mathrm{diss}}$ between the two plates. The numerical results show that both $\mathrm{Im}\Gamma$ and $F_{\mathrm{diss}}$ exhibit a threshold and are positively correlated with $v$. Notably, the dependence of $F_{\mathrm{diss}}$ on $v$ follows a linear relationship, leading to a linear friction coefficient at zero temperature.

It is worth highlighting that we employed a specific physical motivation for the nonlocal interaction $U$ in our calculations. The coupling between the two plates is assumed to be s-d coupling, which originates from fundamental electromagnetic interactions and is inherently nonlocal.
Since condensed matter systems often feature complex and novel excitations, this method could also be applied to other systems beyond electrons. However, for strongly correlated systems without a well-defined quasi-particle picture, non-perturbative approaches to quantum dissipation are also worth exploring.

\section*{Acknowledgement}
\indent We gratefully acknowledge Qiang Sun for valuable discussions. This research was supported by Yunnan Fundamental Research Projects (Grant No. 202401AU070125), the Special Basic Cooperative Research Programs of Yunnan Provincial Undergraduate Universities' Association (Grant No. 202301BA070001-114), Yunnan Provincial Department of Education Science Research Fund Project (No. 2025J0942), 2025 Self-funded Science and Technology Projects of Chuxiong Prefecture (Grant No. cxzc2025004, cxzc2025008), Chuxiong Normal University Doctoral Research Initiation Fund Project (No. BSQD2407, BSQD2507), Yunnan Provincial Department of Education Science Research Fund Project (No. 2025J0942) and Dongying Science Development Fund Project (No. DJB2023015).


\bibliography{bibliography.bib}



\end{document}